\documentclass[aps,twocolumn,amsmath,amssymb,superscriptaddress,prb]{revtex4}
\usepackage{epsf}
\usepackage{graphicx}
\usepackage{gensymb}
\usepackage{sidecap}

\begin{document}

\title{Metallic~surface~electronic~state in~half-Heusler~compounds~\textit{R}PtBi~(\textit{R} = Lu, Dy, Gd)}

\author{Chang~Liu}
\affiliation{Division of Materials Science and Engineering, Ames
Laboratory, Ames, Iowa 50011, USA} \affiliation{Department of
Physics and Astronomy, Iowa State University, Ames, Iowa 50011, USA}

\author{Yongbin~Lee}
\affiliation{Division of Materials Science and Engineering, Ames
Laboratory, Ames, Iowa 50011, USA}

\author{Takeshi~Kondo}
\affiliation{Division of Materials Science and Engineering, Ames
Laboratory, Ames, Iowa 50011, USA} \affiliation{Department of
Physics and Astronomy, Iowa State University, Ames, Iowa 50011, USA}

\author{Eun~Deok~Mun}
\affiliation{Division of Materials Science and Engineering, Ames
Laboratory, Ames, Iowa 50011, USA} \affiliation{Department of
Physics and Astronomy, Iowa State University, Ames, Iowa 50011, USA}

\author{Malinda~Caudle}
\affiliation{Division of Materials Science and Engineering, Ames
Laboratory, Ames, Iowa 50011, USA} \affiliation{Department of
Physics and Astronomy, Iowa State University, Ames, Iowa 50011, USA}

\author{Bruce~N.~Harmon}
\affiliation{Division of Materials Science and Engineering, Ames
Laboratory, Ames, Iowa 50011, USA} \affiliation{Department of
Physics and Astronomy, Iowa State University, Ames, Iowa 50011, USA}

\author{Sergey~L.~Bud'ko}
\affiliation{Division of Materials Science and Engineering, Ames
Laboratory, Ames, Iowa 50011, USA} \affiliation{Department of
Physics and Astronomy, Iowa State University, Ames, Iowa 50011, USA}

\author{Paul~C.~Canfield}
\affiliation{Division of Materials Science and Engineering, Ames
Laboratory, Ames, Iowa 50011, USA} \affiliation{Department of
Physics and Astronomy, Iowa State University, Ames, Iowa 50011, USA}

\author{Adam~Kaminski}
\affiliation{Division of Materials Science and Engineering, Ames
Laboratory, Ames, Iowa 50011, USA} \affiliation{Department of
Physics and Astronomy, Iowa State University, Ames, Iowa 50011, USA}

\date{\today}

\begin{abstract}

Rare-earth platinum bismuth ($R$PtBi) has been recently proposed to
be a potential topological insulator. In this paper we present
measurements of the metallic surface electronic structure in three
members of this family, using angle resolved photoemission
spectroscopy (ARPES). Our data shows clear spin-orbit splitting of
the surface bands and the Kramers' degeneracy of spins at the
$\bar{\Gamma}$ and $\bar{M}$ points, which is nicely reproduced with
our full-potential augmented plane wave calculation for a surface
electronic state. No direct indication of topologically non-trivial
behavior is detected, except for a weak Fermi crossing detected in
close vicinity to the $\bar{\Gamma}$ point, making the total number
of Fermi crossings odd. In the surface band calculation, however,
this crossing is explained by another Kramers' pair where the two
splitting bands are very close to each other. The classification of
this family of materials as topological insulators remains an open
question.

\end{abstract}

\maketitle

\section{Introduction}

The discovery of topologically non-trivial states of matter opens up
a new realm of knowledge for fundamental condensed matter physics.
Unlike conventional materials, these ``topological insulators"
exhibit metallic surface states that are protected by time reversal
symmetry, while maintaining an insulating bulk electronic structure.
This leads to a variety of novel properties including odd number of
surface Dirac fermions, strict prohibition of back-scattering, etc.,
paving the way to potential technical breakthroughs in e.g. quantum
computing process via the application of
spintronics\cite{Hasan_review, Moore}. Recently, extensive
theoretical and experimental efforts have led to the realization of
such fascinating behaviors in e.g. the HgTe quantum
wells\cite{Zhang_HgTe, Zhang_HgTe2, Zhang_HgTe3}, the
Bi$_{1-x}$Sb$_x$ system\cite{Hsieh_BiSb, Hsieh_BiSb2, Yazdani_BiSb}
and the Bi$_2$X$_3$ (X = Te, Se) binary compounds\cite{Zhang_Bi2Se3,
Shen_Bi2Te3}. Numerous half-Heusler ternary compounds have been
proposed, theoretically, to be potential new platforms for
topological quantum phenomena\cite{Zhang_Heusler, Hasan_Heusler},
where the inherent flexibility of crystallographic, electronic and
superconducting parameters provide a multidimensional basis for both
scientific and technical exploration. The experimental determination
of their topological class would set the basis for possible
spintronic utilization and further studies on the interplay between
the topological quantum phenomena versus e.g. the
magnetic\cite{Canfield}, superconducting\cite{Goll} and heavy
Fermionic\cite{Fisk} behaviors.

Theoretically, the topological insulators experience a gapless
surface state protected by time reversal symmetry and thus are
robust against scattering from local impurities. Such a surface
state is ``one half" of a normal metal in that the surface bands are
strongly spin-polarized, forming a unique spin helical
texture\cite{Hsieh_BiSb2, Hsieh_Bi2Se3}. On the other hand, the
Kramers' theorem requires that the spin be degenerate at the
Kramers' points - $k$-points of the surface Brillouin zone where
time reversal symmetry is preserved\cite{Kane_Z2}. At the interface
between, say, a normal spin-orbit system and vacuum, the
spin-polarized surface bands connect pairwise (Kramers' pair),
crossing the chemical potential $\mu$ an even number of times
between two distinct Kramers' points. At the interface between a
topologically non-trivial material and vacuum, however, one expects
the surface bands to cross $\mu$ an odd number of
times\cite{Hasan_review}.

In this paper we present a systematic survey on the surface
electronic structure of half-Heusler compounds \textit{R}PtBi
(\textit{R} = Lu, Dy, Gd) using angle resolved photoemission
spectroscopy (ARPES). Our results show clear spin-orbit splitting of
the surface bands that cross the chemical potential, which is nicely
reproduced in the full-potential augmented plane wave calculation
for a surface electronic state. The Kramers' degeneracy of spin is
unambiguously detected at both the $\bar{\Gamma}$ and $\bar{M}$
points. No direct indication of topologically non-trivial behavior
is detected, except for the fact that there is a weak Fermi crossing
in the close vicinity to the $\bar{\Gamma}$ point, making the total
number of crossings five. In the surface band calculation, however,
this inner crossing is explained by two spin-orbit splitting bands
that are very close to each other, forming another Kramers' pair. In
this band configuration, the total Berry phase would be zero for the
half-Heusler systems, and they would not be topologically
non-trivial. The detailed topological class of this family of
materials thus remains an open question, requiring a detailed
spin-resolved ARPES study with ultra-high momentum resolution and a
direct calculation of the topological invariants based on the first
principle band structure.

\begin{figure*}[t]
\includegraphics[width=7.1in]{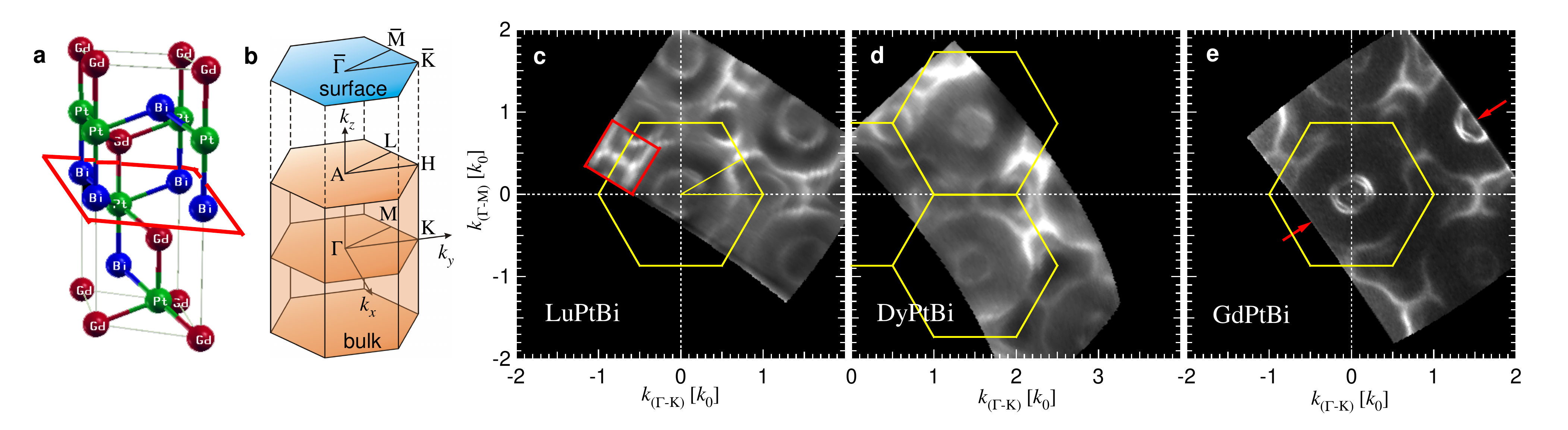}
\caption{(Color online) Surface Fermi maps of half-Heusler compounds
\textit{R}PtBi (\textit{R} = Lu, Dy, Gd). (a) $C1_b$ Crystal
structure of \textit{R}PtBi. The crystallographic axes are rotated
so that the (111) direction points along $z$. The red parallelogram
marks the Bi(111) cleaving plane. (b) The surface and bulk Brillouin
zone for the rotated crystal structure in (a). Here $k_z$
corresponds to the (111) direction of the \textit{fcc} Brillouin
zone. (c)-(e) Surface Fermi maps of \textit{R}PtBi. All data is
taken with 48 eV photons at $T=15$ K. Yellow lines denote the
surface Brillouin zone.} \label{Fig1}
\end{figure*}

\begin{figure}[h!]
\includegraphics[width=3.2in]{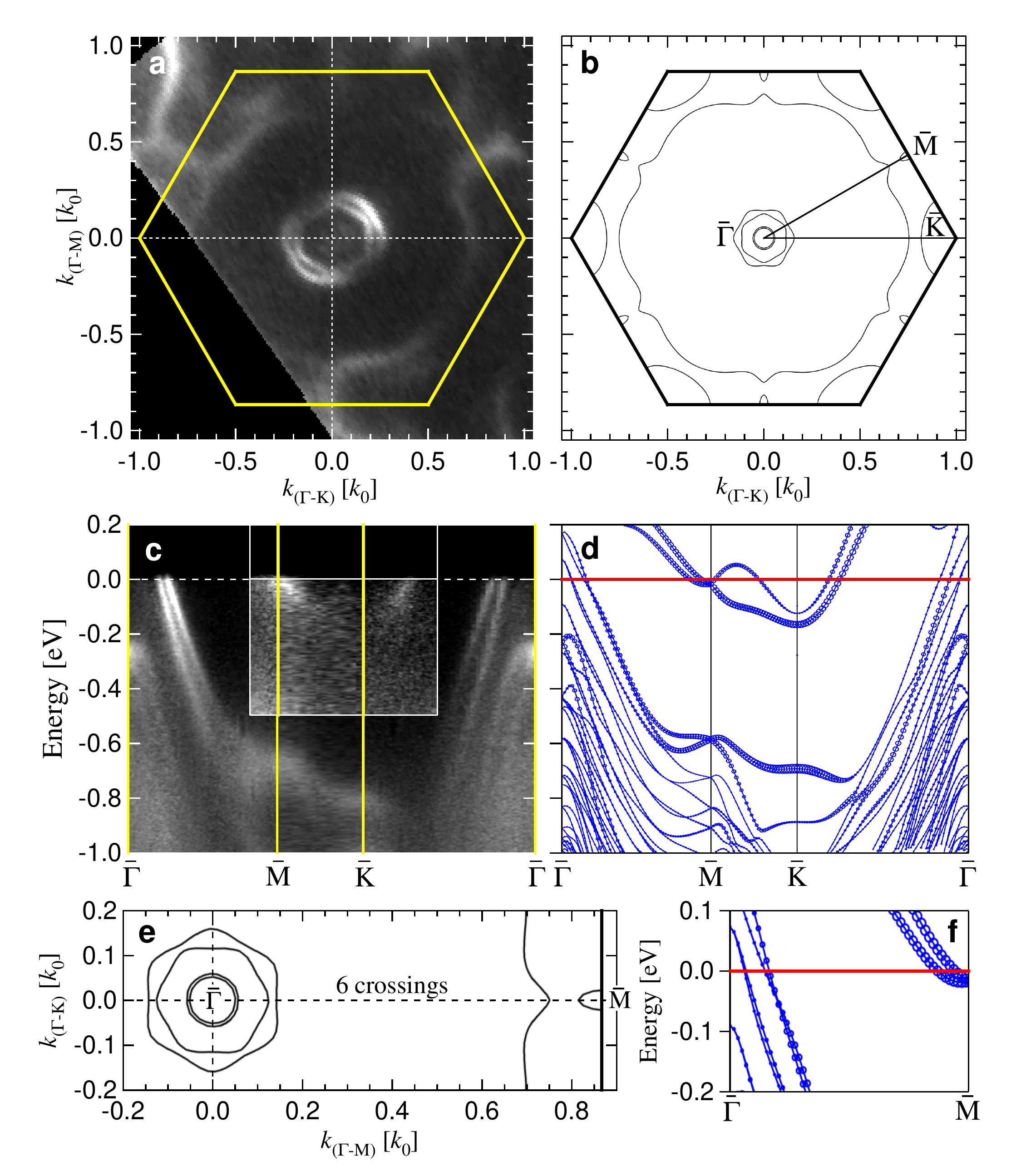}
\caption{(Color online) Surface electronic structure of GdPtBi:
Comparison between ARPES data and calculational result. (a) Fermi
map of GdPtBi observed by ARPES, same as Fig. 1(e). (b)
Calculational surface Fermi map of GdPtBi at the Bi(111) cleaving
plane. See text for details. (c) ARPES band structure along the
contour $\bar{\Gamma}$-$\bar{M}$-$\bar{K}$-$\bar{\Gamma}$. Inset of
(c) enhanced the ARPES intensity near $\bar{M}$ and $\bar{K}$ for
better visibility of the bands. (d) Calculational band structure
with respect to (c). (e)-(f) Expanded figures for (b) and (d),
respectively, showing six Fermi crossings. Panel (e) is rotated by
30$^\circ$ with respect to (b).} \label{Fig2}
\end{figure}

\begin{SCfigure*}
\centering
\includegraphics[width=4.5in]{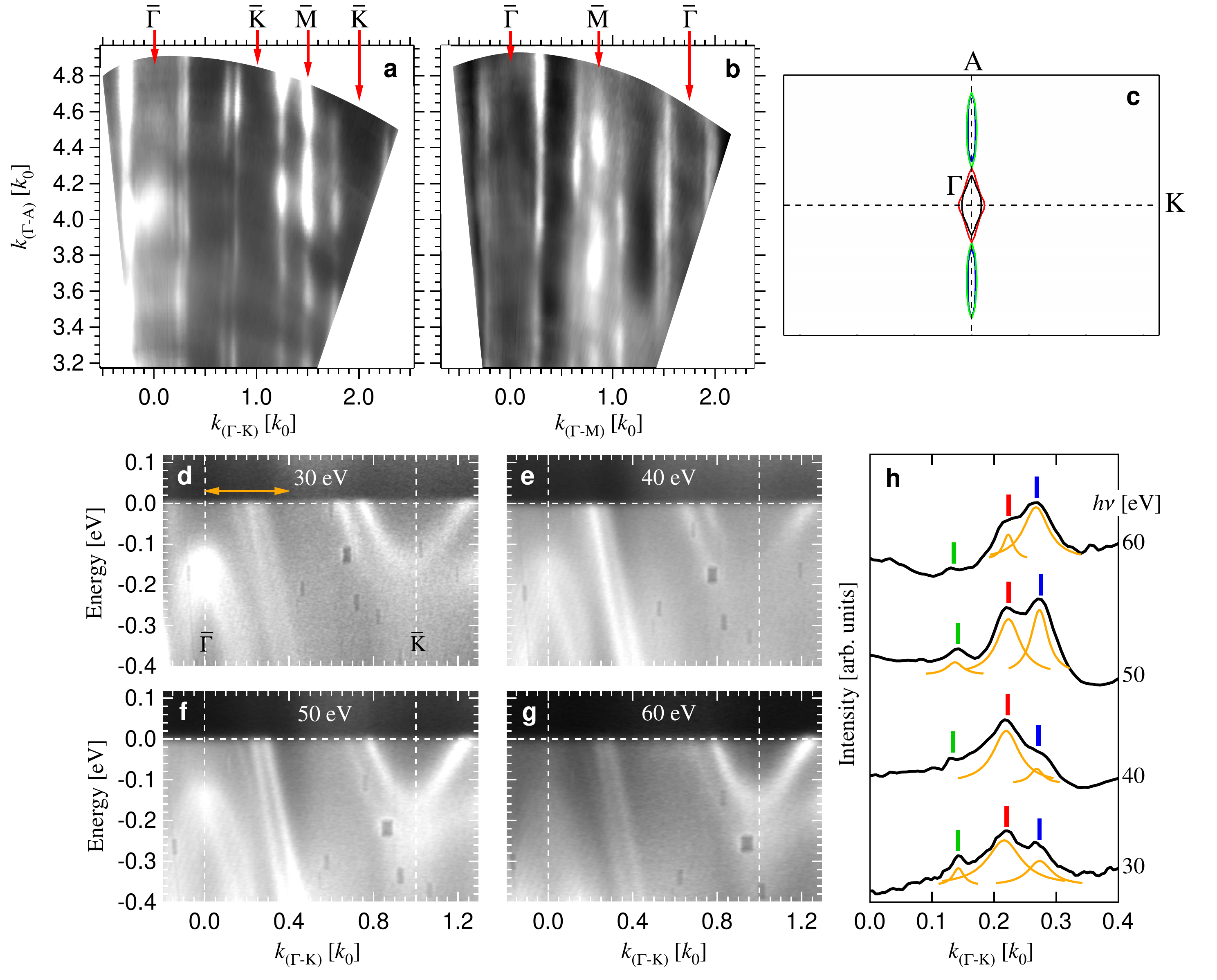}
\caption{(Color online) Absence of $k_z$ dispersion as proof for the
observation of a surface electronic structure. (a)-(b) $k_z$
($\Gamma$-$A$) dispersion maps for LuPtBi. Data is obtained by
scanning the incident photon energy $h\nu$ from 30 to 80 eV along
(a), the $\bar{\Gamma}$-$\bar{K}$ and (b) the
$\bar{\Gamma}$-$\bar{M}$ direction. (c) Calculational Fermi surface
map for the \textit{bulk} state in the $A$-$\Gamma$-$K$ plane. See
panel (a) for comparison. (d)-(g) Band dispersion maps along the the
$\bar{\Gamma}$-$\bar{K}$ direction for selected $h\nu$s. It is clear
that all observed bands are independent of $h\nu$ ($k_z$). (h)
Detailed peak analysis for the momentum distribution curves (MDCs)
at the chemical potential for four different photon energies. $k$
range is indicated by an orange double arrow in (d). Bars in
different colors indicate the Fermi crossings for different bands.}
\label{Fig3}
\end{SCfigure*}

\section{Experimental}

Single crystals of \textit{R}PtBi (\textit{R} = Lu, Dy, Gd) were
grown out of a Bi flux and characterized by room temperature power
X-ray diffraction measurements\cite{Canfield, Growth}. The crystals
grow as partial octahedra with the (111) facets exposed. Typical
dimensions of a single crystal are about $0.5\times0.5\times0.5$
$\textrm{mm}^3$. The ARPES measurements were performed at beamline
10.0.1 of the Advanced Light Source (ALS), Berkeley, California
using a Scienta R4000 electron analyzer. Vacuum conditions were
better than $3\times10^{-11}$ torr. All ARPES data was taken at
$T=15$ K, above the magnetic ordering temperatures of all
compounds\cite{Canfield}. The energy resolution was set at $\sim$ 15
meV. All samples were cleaved \textit{in situ}, yielding clean (111)
surfaces in which atoms arrange in a hexagonal lattice. High
symmetry points for the surface Brillouin zone are defined as
$\bar{\Gamma}(0,0)$, $\bar{K}(k_0,0)$ and $\bar{M}(0,
k_0\sqrt{3}/2)$ with unit momentum $k_0=\sqrt{6}\pi/a$, where $a$ is
the lattice constant for each type of crystals. We emphasize here
that no stress or pulling force is felt by the samples, which
ensures that the measured data reveals the intrinsic electronic
structure of the single crystals.

\section{Results and discussion}

We begin this survey in Fig. 1 by showing the Fermi maps of the
three half-Heusler compounds \textit{R}PtBi (\textit{R} = Lu, Dy,
Gd). Previous theoretical calculations for the bulk electronic
structure\cite{Zhang_Heusler, Hasan_Heusler, Antonov} suggested that
the Kramers' crossing at the $\bar{\Gamma}$ point happens very close
to $\mu$; the Fermi surface reduces to a single point (Dirac point)
at $\bar{\Gamma}$. The data in Fig. 1 shows that, at least in the
(111) cleaving plane, this is not the case. Instead there are
several bands crossing $\mu$ in the vicinity of both the
$\bar{\Gamma}$ and $\bar{M}$ points. The overall Fermi surface for
all three half-Heusler compounds are similar, indicating a similar
cleaving plane and band structure for all members. By comparing the
band structure measured at the (111) surface with results of band
calculations for GdPtBi (Fig. 2), we find the cleaving plane to be
Bi(111), marked by a red parallelogram in Fig. 1(a). A closer look
at Fig. 1(c)-(e) reveals that the $\bar{\Gamma}$ pockets have
different sizes for different Heusler members. For example the
circular $\bar{\Gamma}$ pockets in LuPtBi are larger in size than
those in GdPtBi. This indicates a different effective electron
occupancy for different members of the half-Heusler family. One
should also note that in Fig. 1(e) the inner of the two bright
$\bar{\Gamma}$ pockets is hexagonal in shape, reminiscent of the
hexagonal shape of the Dirac cone in Bi$_2$Te$_3$ (Ref.
\cite{Shen_Bi2Te3}), which is explained by higher order terms in the
$k \cdot p$ Hamiltonian\cite{Fu}. This hexagonal shape is very
nicely reproduced in the calculation [Fig. 2(b)]. For clarifying the
topological class of the half-Heuslers, two immediate questions
follow the observations in Fig. 1: (1) Are the observed bands
actually arising due to the sample surface? (2) Exactly how many
times do the bands intersect the chemical potential along the
$\bar{\Gamma}$-$\bar{M}$ line segment?

\begin{SCfigure*}
\centering
\includegraphics[width=4.8in]{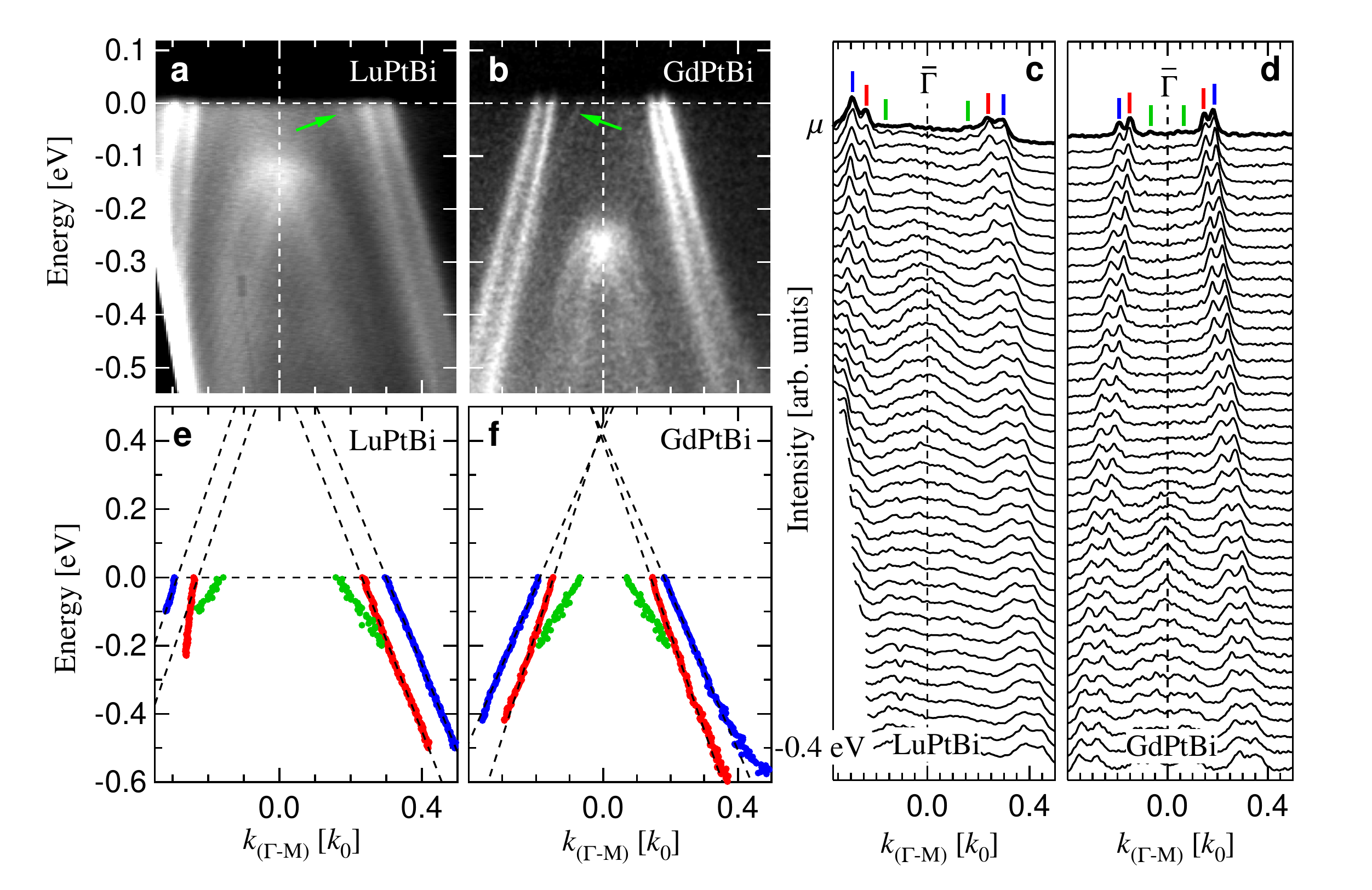}
\caption{Band structure analysis at the vicinity of $\bar{\Gamma}$
[red arrows in Fig. 1(e)]. Data is taken on LuPtBi and GdPtBi
samples at $T=15$ K. (a)-(b) Band dispersion maps along the
$\bar{\Gamma}$-$\bar{M}$ direction. Green arrows point to the
position of the inner hole band which have lower intensity than the
two other hole bands. (c)-(d) Corresponding MDCs for panels (a) and
(b). (e)-(f) Extraction of the band position for panels (a) and (b).
By linearly extrapolating the bands above the chemical potential
$\mu$ we show an approximate band crossing point (Dirac point) at
$E\sim0.4$ eV for GdPtBi.}
\end{SCfigure*}

\begin{figure}[t]
\includegraphics[width=3.5in]{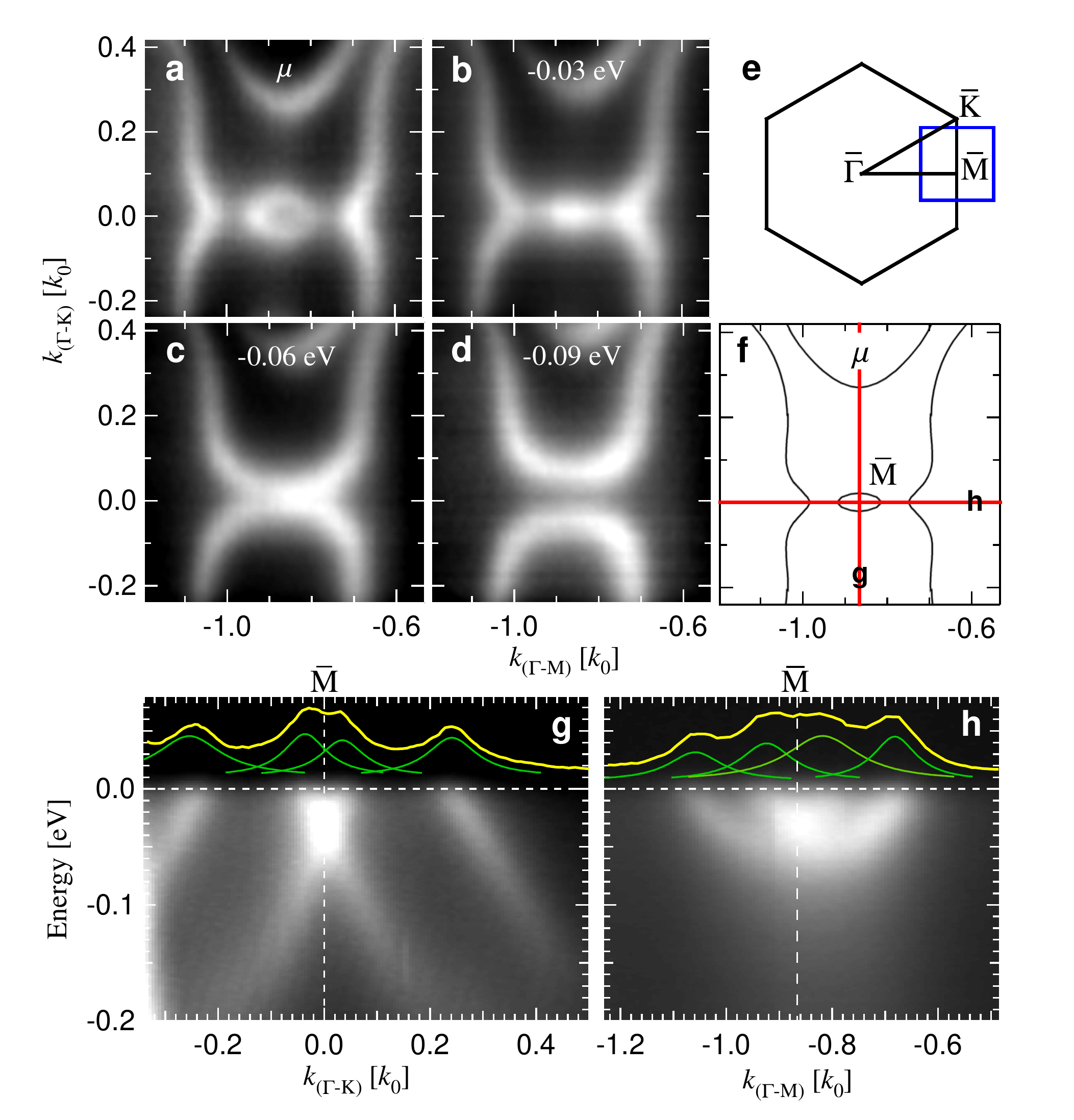}
\caption{Band structure analysis at the vicinity of $\bar{M}$ [red
box in Fig. \ref{Fig1}(c)]. Data is taken on LuPtBi samples. (a)-(d)
Binding energy dependence of band structure near $\bar{M}$. Map
location in the surface Brillouin zone is shown in Panel (e). (f)
Theoretical band map at the chemical potential for GdPtBi. (g),(h)
Band maps for two perpendicular directions marked by red lines in
(g). There are in total two Fermi crossings along the
$\bar{\Gamma}$-$\bar{M}$ line segment at the vicinity of $\bar{M}$.}
\label{Fig5}
\end{figure}

Fig. 2 shows the comparison between the ARPES data and a
calculational surface state in GdPtBi. For both the band structure
and Fermi surface calculation, we used a full-potential linear
augmented plane wave (FPLAPW) method\cite{wien2k} with a local
density functional\cite{LDA}. The crystallographic unit cell is
generated such that the (111) direction of the \textit{fcc}
Brillouin zone points along the $z$-axis. For calculation of the
surface electronic structure, supercells with three unit cell layers
and a 21.87 a.u. vacuum is constructed. We calculated band
structures of all six possible surface endings (Gd-Bi-Pt-bulk,
Gd-Pt-Bi-bulk, Bi-Gd-Pt-bulk, Bi-Pt-Gd-bulk, Pt-Gd-Bi-bulk,
Pt-Bi-Gd-bulk); only the Bi-Pt-Gd-bulk construction shows good
agreement with experiment [Fig. 2(b), (d)-(f)]. Structural data were
taken from a reported experimental result\cite{Haase}. To obtain the
self-consistent charge density, we chose 48 $k$-points in the
irreducible Brillouin zone, and set $R_{\textrm{MT}}\times
k_{\textrm{max}}$ to 7.5, where $R_{\textrm{MT}}$ is the smallest
muffin-tin radius and $k_{\textrm{max}}$ is the plane-wave cutoff.
We used muffin-tin radii of 2.5, 2.4 and 2.4 a.u. for Gd, Bi, and Pt
respectively. For the non-magnetic-state calculation valid for
comparison with ARPES results at 15 K, the seven 4\textit{f}
electrons of Gd atoms were treated as core electrons with no net
spin polarization. The atoms near the surface (Bi, Pt, Gd) were
relaxed along the $z$-direction until the forces exerted on the
atoms were less than 2.0 mRy/a.u.. With this optimized structure, we
obtained self-consistency with 0.01 mRy/cell total energy
convergence. After that, we calculated the band structure and two
dimensional Fermi surface in which we divided the rectangular cell
connecting four $\bar{K}$-points by $40 \times 40$, yielding 1681
$k$-points.

Even at first glance, Fig. 2 gives the impression of remarkable
agreement between theory and experiment. All basic features observed
by ARPES - the overall shape and location of the Fermi pockets [Fig.
2(a)-(b)], the binding energies of the bands [Fig. 2(c)-(d)] - are
well reproduced by the calculation. The main point of this figure,
however, is the fact that band calculations show a total of six
Fermi crossings along the $\bar{\Gamma}$-$\bar{M}$ line segment,
which is an even number and is not directly consistent of the
proposed strong topological insulating
phenomenon\cite{Hasan_Heusler, Zhang_Heusler}. It should be noted
that, in order to take into account the spin-orbit splitting,
relativistic effects are applied to the calculation. Similar
calculations reproduce clear topological insulating behavior in
Bi$_2$Te$_3$ thin films\cite{Park}. The excellent agreement shown in
Fig. 2 also implies the validity of such calculation in half-Heusler
compounds. In fact traces for the inner two crossings is also found
in the ARPES data, where they appear to be one single crossing, most
likely due to finite momentum resolution [Leftmost part in Fig.
2(c), see also Fig. 3(d)-(h)].

In Fig. 3 we prove that the observed bands come from the sample
surface. This is done by scanning the incident photon energy along
both $\bar{\Gamma}$-$\bar{K}$ and $\bar{\Gamma}$-$\bar{M}$ high
symmetry directions. Varying the photon energy in ARPES effectively
changes the momentum offset along the direction perpendicular to the
sample surface. In our case, this direction corresponds to $k_z$ or
the (111) direction of the \textit{fcc} Brillouin zone. Figs.
3(a)-(b) show that all observed bands form straight lines along the
$k_z$ direction, a clear indication for the lack of $k_z$
dependence. In Fig. 3(c) we compare this to a calculated Fermi
surface map for the \textit{bulk} bands, along the same direction as
in Fig. 3(a). The difference is clear: the bulk bands are dispersive
along the $\Gamma$-$A$ direction; and most of the experimentally
observed bands are not present in the calculation. In Figs. 3(d)-(h)
we pay special attention to the bands crossing $\mu$ near
$\bar{\Gamma}$ by showing the band structure for four different
photon energies. In total there are at least three Fermi contours
surrounding $\bar{\Gamma}$, the outer two being a lot brighter than
the inner one (or two, see discussion for Fig. 2). As shown in Fig.
3(h), These three (or four) bands cross $\mu$ at exactly the same
$k$ positions for all photon energies. Therefore all of them are
surface bands. The data in Fig. 3 thus show, unambiguously, that a
metallic surface electronic state exists in the half-Heusler
compounds.

The exact number of Fermi crossings along the
$\bar{\Gamma}$-$\bar{M}$ line segment is also examined in Fig. 4.
The main conclusion for Fig. 4 is that there are also three (or
four) visible Fermi crossings at the vicinity of $\bar{\Gamma}$
between these two Kramers' points. We show these bands on the LuPtBi
and GdPtBi samples. Both on the band dispersion maps [Figs.
4(a)-(b)] and the momentum distribution curves [MDCs, Figs.
4(c)-(d)] we see that there are two bright hole-like bands almost
parallel to each other, and a much weaker inner band with lower
Fermi velocity. This inner band is not easy to see in the band maps
(nontheless indicated by green arrows), but is clearly visible in
the MDCs by small intensity peaks tracing down from the one marked
by a green bar [also marked by a green color in Figs. 4(e)-(f)]. The
same band also exists in the $\bar{\Gamma}$-$\bar{K}$ direction
[Figs. 3(d)-(h)]. Same as the discussion for Figs. 2 and 3, this
inner crossing is reproduced in the band calculation by two closely
located spin-orbit-splitting bands that form a Kramers' pair. The
brighter parallel bands form a second Kramers' pair of opposite
spins. In Fig. 4(e)-(f) we show the linear extrapolation of the two
brighter bands. In GdPtBi they are likely to reduce to a Dirac point
at about 0.4 eV above $\mu$. If the total number of crossing is
four, such a configuration will give zero contribution to the total
Berry phase.

In Fig. 5 we examine the bands near the $\bar{M}$ point. The
$k$-space location of the ARPES maps [Figs. 5(a)-(d)] is shown in
Fig. 5(e). Panels 5(g)-(h) present the band dispersion maps for two
cuts crossing $\bar{M}$, whose positions are marked in Panel 5(f)
with the band calculation result. Figs. 5(a)-(d) show that the
$\bar{M}$ bands form a very special shape. At high binding energies
[$E\sim-0.1$ eV, Fig. 5(d)], two U-shape bands are well separated.
As binding energy decreases these two bands merge into each other
and hybridize to form a central elliptical contour and two
curly-bracket-like segments. The segments near each $\bar{M}$ points
link together, forming another large Fermi contour enclosing the
zone center $\bar{\Gamma}$. It is clear from Fig. 5(g)-(h) that
there are two Fermi crossings in both the $\bar{\Gamma}$-$\bar{K}$
and $\bar{\Gamma}$-$\bar{M}$ directions. The special shape of the
Fermi surface is formed by two bands that are likely to be members
of another Kramers' pair. Kramers' degeneracy of spin happens at
$\sim30$ meV below $\mu$. All this features are obtained with our
calculation for the surface states [Fig 2(b) and 2(e)]. These two
bands also give zero contribution to the total Berry phase.

In summary, we performed an ARPES survey on the electronic structure
of three half-Heusler compounds \textit{R}PtBi (\textit{R} = Lu, Dy,
Gd) which are proposed to be topological insulators. Our result show
unambiguously that these materials have a metallic surface state
markedly different from the calculational result on the bulk
electronic structures. This surface state is reproduced with high
accuracy in our band calculations. Both experiment and theory reveal
several bands that cross the Fermi level. Knowledge of the exact
number of these bands is possibly limited by experimental momentum
resolution. No direct consistency with the proposed strong
topological insulating behavior is found in the ARPES results. For
final determination of their topological classes, both an APRES
measurement of ultrahigh $k$-resolution and a direct calculation of
the first Chern number as a topological invariant \cite{Qi} are in
need.

\section{Acknowledgement}

We thank S.-C. Zhang and J. Schmalian for instructive discussions as
well as Sung-Kwan Mo for grateful instrumental support at the ALS.
Ames Laboratory was supported by the Department of Energy - Basic
Energy Sciences under Contract No. DE-AC02-07CH11358. ALS is
operated by the US DOE under Contract No. DE-AC03-76SF00098.

\end{document}